\documentclass[]{zakoproc}
\usepackage{graphicx}
\begin{document}

\title{The LOFAR Magnetism Key Science Project}
\author{James Anderson$^{1}$
        Rainer Beck$^{1}$
        Michael Bell$^{2}$
        Ger de Bruyn$^{3,4}$\\
        Krzysztof Chy\.{z}y$^{5}$
        Jochen Eisl\"offel$^{6}$
        Torsten En{\ss}lin$^{2}$
        Andrew Fletcher$^{7}$\\
        Marijke Haverkorn$^{8}$
        George Heald$^{2}$
        Andreas Horneffer$^{1}$
        Aris Noutsos$^{1}$\\
        Wolfgang Reich$^{1}$
        Anna Scaife$^{9}$
        -- on behalf of the LOFAR collaboration}

\institute{$^{1}$ Max-Planck-Institut f\"ur Radioastronomie,
           Auf dem H\"ugel 69, 53121 Bonn, Germany\\
           $^{2}$ Max-Planck-Institut f\"ur Astrophysik, Karl-Schwarzschild-Str.1, 85741 Garching, Germany\\
           $^{3}$ ASTRON, PO Box 2, 7990 AA Dwingeloo, The Netherlands\\
           $^{4}$ Kapteyn Institute, PO Box 800, 9700 AV Groningen,
           The Netherlands\\
           $^{5}$ Astronomical Observatory, Jagiellonian University,
           ul. Orla 171, 30-244 Krak\'ow, Poland\\
           $^{6}$ Th\"uringer Landessternwarte Tautenburg, Sternwarte 5, 07778 Tautenburg, Germany\\
           $^{7}$ School of Mathematics and Statistics,
           Newcastle University, Newcastle NE1 7RU, UK\\
           $^{8}$ Radboud University Nijmegen, PO Box 9010, 6500 GL Nijmegen, The Netherlands\\
           $^{9}$ University of Southampton, Highfield, Southampton SO17 1BJ, UK
           }

\markboth{J. Anderson et al.}{The LOFAR Magnetism Key Science
Project}

\maketitle

\begin{abstract}

Measuring radio waves at low frequencies offers a new window to
study cosmic magnetism, and LOFAR is the ideal radio telescope to
open this window widely. The LOFAR Magnetism Key Science Project
(MKSP) draws together expertise from multiple fields of magnetism
science and intends to use LOFAR to tackle fundamental questions on
cosmic magnetism by exploiting a variety of observational
techniques. Surveys will provide diffuse emission from the Milky Way
and from nearby galaxies, tracking the propagation of long-lived
cosmic-ray electrons through magnetic field structures, to search
for radio halos around spiral and dwarf galaxies and for magnetic
fields in intergalactic space. Targeted deep-field observations of
selected nearby galaxies and suspected intergalactic filaments allow
sensitive mapping of weak magnetic fields through Rotation Measure
(RM) grids. High-resolution observations of protostellar jets and
giant radio galaxies reveal structures on small physical scales and
at high redshifts, whilst pulsar RMs map large-scale magnetic
structures of the Galactic disk and halo in revolutionary detail.
The MKSP is responsible for the development of polarization
calibration and processing, thus widening the scientific power of
LOFAR.

\end{abstract}

\section{Introduction}

Magnetic fields are pervasive throughout the Universe. Obtaining a
detailed knowledge of the strength, morphology and evolution of
these magnetic fields is essential not only for understanding the
energetics and dynamics of numerous astrophysical phenomena, but
using cosmic large scale structure as a laboratory in which to probe
these fields is also the key to unlocking the even more fundamental
long standing problems of magnetic evolution and structure, and
ultimately to determining the origin of cosmic magnetism itself.

At radio wavelengths, magnetic fields reveal themselves indirectly
in two major ways: synchrotron emission from cosmic-ray electrons
(CRe) and Faraday rotation of polarized background emission due to
magnetized media along the line of sight. The synchrotron radiation
traces the component of the total field in the sky plane, while the
Faraday rotation measure (RM) provides information on the field
component along the line of sight (Sect.~5). Both of these
mechanisms are enhanced at low radio frequencies due to the
increased intensity of synchrotron emission and the
wavelength-squared dependence of Faraday rotation, making LOFAR
uniquely suited to magnetism science. The accuracy in RM
measurements increases with increasing total (not necessarily
contiguous) coverage in wavelength space (Brentjens \& de Bruyn
\cite{brentjens05}, Heald \cite{heald09}), so that much smaller
variations in field strength can be measured at low frequencies. At
such low frequencies CRe have longer lifetimes, traveling further
from their site of generation into distant regions of low magnetic
field strength and revealing rarified magnetic structures invisible
at higher frequencies. These structures carry the signature of
astrophysical processes that have been erased in more active
regions, but which remain crucial for our understanding of cosmic
history. Galaxies are expected to be much larger than their optical
sizes, when observed at low radio frequencies.

A brief summary of the imaging capabilities of LOFAR was given by Heald et
al. (\cite{heald11}).

\section{Science areas}

The Magnetism Key Science Project (MKSP) aims to exploit the unique
abilities of LOFAR to investigate cosmic magnetic fields in a
variety of astrophysical sources. The MKSP Project Plan includes an
initial target list of galaxies, selected in close cooperation with
the LOFAR Surveys Key Science Project (R\"ottgering et al.
\cite{rottgering_2011}), followed by deep observations of galaxies
and galaxy groups. These deep fields will also serve as targets to
investigate magnetic fields in the Milky Way foreground. The
structure of small-scale magnetic fields will be studied the lobes
of giant radio galaxies. Polarized synchrotron emission and rotation
measures from pulsars and polarized jets from young stars will be
observed in cooperation with the Transients Key Science Project
(Fender et al. \cite{fender06}, Stappers et al. \cite{stappers11}).

\subsection{Milky Way}

LOFAR's broad coverage of low frequencies makes it uniquely suited
to studying weak magnetic fields and low-density regions in the halo
of the Milky Way, where such studies can address both the disk-halo
interaction and interstellar medium (ISM) energetics. In addition to
probing the origin and evolution of Galactic magnetism through the
regular field component and interstellar turbulence though
small-scale structure, Galactic halo magnetic field investigations
also importantly address the confusing effect of Galactic emission
as a Faraday screen when studying distant extended galactic and
extragalactic objects: at LOFAR frequencies foreground influence is
expected in all directions and must be properly separated for a
correct interpretation of extragalactic data. High-resolution 3-D
simulations of Galactic emission components will be used for this
purpose (Sun et al. \cite{sun09}). Rotation Measure (RM) Synthesis
of diffuse Galactic synchrotron emission is an excellent way to
disentangle various RM components, giving statistical information on
the clumped magnetized ISM and on the relation of thermal electron
density to magnetic field strength.

LOFAR will also be an excellent tracer of the magnetized ISM due to
its high angular resolution, which minimizes beam depolarization
effects, and due to its coverage at low frequencies enabling it to
detect aged electrons high in the Galactic halo that emit at low
frequencies. Within a few kpc the 3-D structure of the Galactic
magnetic field will be mapped with unprecedented accuracy,
complementary to pulsar RMs (Sect.~2.2). Comparison of RM data to
magnetic field models will shed light on the configuration of
large-scale regular fields. The deep fields of nearby galaxies
(Sect.~2.3) with a minimum of Galactic foreground contamination will
be used to make a statistical comparison of RMs and redshifts in
order to investigate the presence and properties of magnetic fields
in the Galactic halo.


The targets of observation for the deep observations of galaxies
(Sect.~2.3) will lie at intermediate and high Galactic latitudes and
will also be used to probe the magnetized Galactic halo. Here, less
is known about the Galactic magnetic field, even though this field
is important in the disk-halo energy exchange and for the
propagation and scattering of extragalactic ultrahigh-energy cosmic
rays (UHECRs).

\subsection{Pulsars}

\begin{figure}
\centering
\includegraphics[width=10cm]{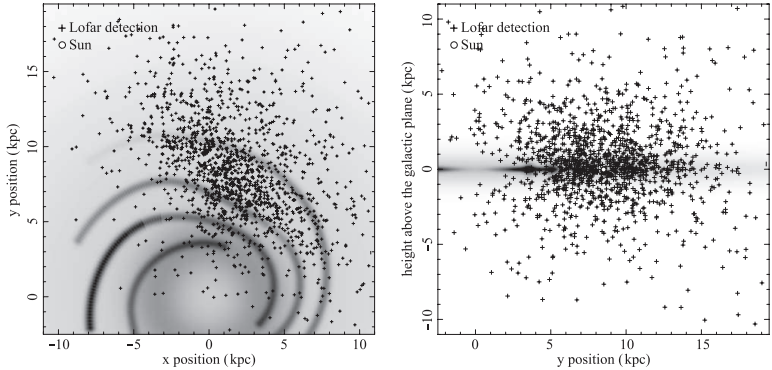}
\caption{Simulation of the 1000+ pulsars that LOFAR is expected to
find in a 60-day all-sky survey, shown in a Galactic plane
projection (left) and an edge-on view of the Galactic plane (right)
(from van Leeuwen \& Stappers \cite{leeuwen10}). }
\label{fig:pulsars}
\end{figure}

Pulsars play a central role in detecting interstellar magnetic
fields in the Galaxy. Polarization observations of hundreds of
pulsars have already been used to map large-scale features in the
Galactic magnetic field (e.g.~Noutsos et al. \cite{noutsos08}): the
field's magnitude and direction along the line of sight to each
pulsar can be determined by using the pulsar RMs. One large-scale
field reversal was found, possibly more exist but cannot be
confirmed with the present data. LOFAR pulsar searches will benefit
from both high sensitivity and an increasing pulsar brightness at
low frequencies. This is expected to result in the discovery of a
new population of dim, nearby and high-latitude pulsars too weak to
be found at higher frequencies: roughly 1,000 pulsar discoveries are
expected from LOFAR (Fig.~\ref{fig:pulsars}). Polarization
observations of these pulsars will approximately double the current
RM sample ($\approx 700$ RMs). When combined with the catalogue of
$\approx 38,000$ extragalactic-source RMs (Taylor et al.
\cite{taylor09}), this will provide the strength and direction of
the regular magnetic field in previously unexplored directions and
locations in the Galaxy; e.g.~very little is known about the
magnetic field properties of the Milky Way beyond a few hundred
parsecs from the Galactic plane. RMs of high-latitude pulsars and
extragalactic sources are crucial for determining fundamental
properties such as the scale height and geometry of the magnetic
field in the thick disk and halo, as well as providing the exciting
prospect of discovering magnetic fields in globular clusters.

In addition to providing an indirect probe of Galactic magnetic
structure, polarization surveys of pulsars at low frequencies will
provide a new view of intrinsic pulsar physics. These include the
effects of scattering, which are prominent at LOFAR frequencies but
have until now only been studied at higher frequencies (e.g. Noutsos
et~al. \cite{noutsos09}); pulsar polarization spectra, which are
expected to turn-over in the LOFAR band ($100-400$\,MHz), and
constraints on the geometry of pulsar magnetospheric emission. We
intend to observe every pulsar detected with LOFAR in polarization
using both LBA and HBA.

\subsection{Nearby galaxies}

It is now generally accepted that galactic magnetic fields result
from the amplification of a seed magnetic field by a hydromagnetic
dynamo, rather than having a merely primordial origin. Turbulent
``seed'' fields in young galaxies can originate from the Weibel
instability in shocks during the cosmological structure formation
(Lazar et al. \cite{lazar09}), injected by the first stars or jets
generated by the first black holes (Rees \cite{rees05}). The
turbulent dynamo can further amplify the field on small timescales
(Schleicher et al. \cite{schleicher10}), followed by the large-scale
dynamo (Beck et al. \cite{beck96}, Arshakian et al.
\cite{arshakian09}).


Important questions still remain regarding the amplification
process, as well as the configuration of the large-scale field in
evolving galaxies (e.g. Moss et~al. \cite{moss12}) and the field
structure in galactic halos (Braun et al. \cite{braun10}). The
expected number density of background sources seen by LOFAR will for
the first time enable systematic studies of galactic field
structures using Faraday rotation of background sources (Stepanov et
al. 2008). ``Faraday spectra'' generated by RM Synthesis (Sect.~5)
will allow a detailed 3-D view of regular magnetic fields and their
reversals (Bell et al. \cite{bell11a}, Frick et al. \cite{frick11})
and enable a clear measurement of magneto-ionic turbulent
fluctuations and their scale spectrum. These will give us a handle
on the properties of the turbulent motions responsible for dynamo
action, allowing us to address outstanding key questions: such as
whether magnetic fields are dynamically important in the ISM of
galaxies at different evolutionary stages. LOFAR's sensitivity to
regions of low density and weak field strengths will also allow us
to measure the magnetic structure in the outer disks and wider halos
of spiral galaxies. It is here that star formation activity is low,
and processes additional to dynamo action, such as gas outflows from
the inner disk, the magneto-rotational instability, gravitational
interaction and ram pressure by the intergalactic medium are
imprinted on this magnetic structure.

\begin{figure*}[t]
\vspace*{2mm}
\begin{minipage}[t]{6.5cm}
\begin{center}
\includegraphics[width=6cm]{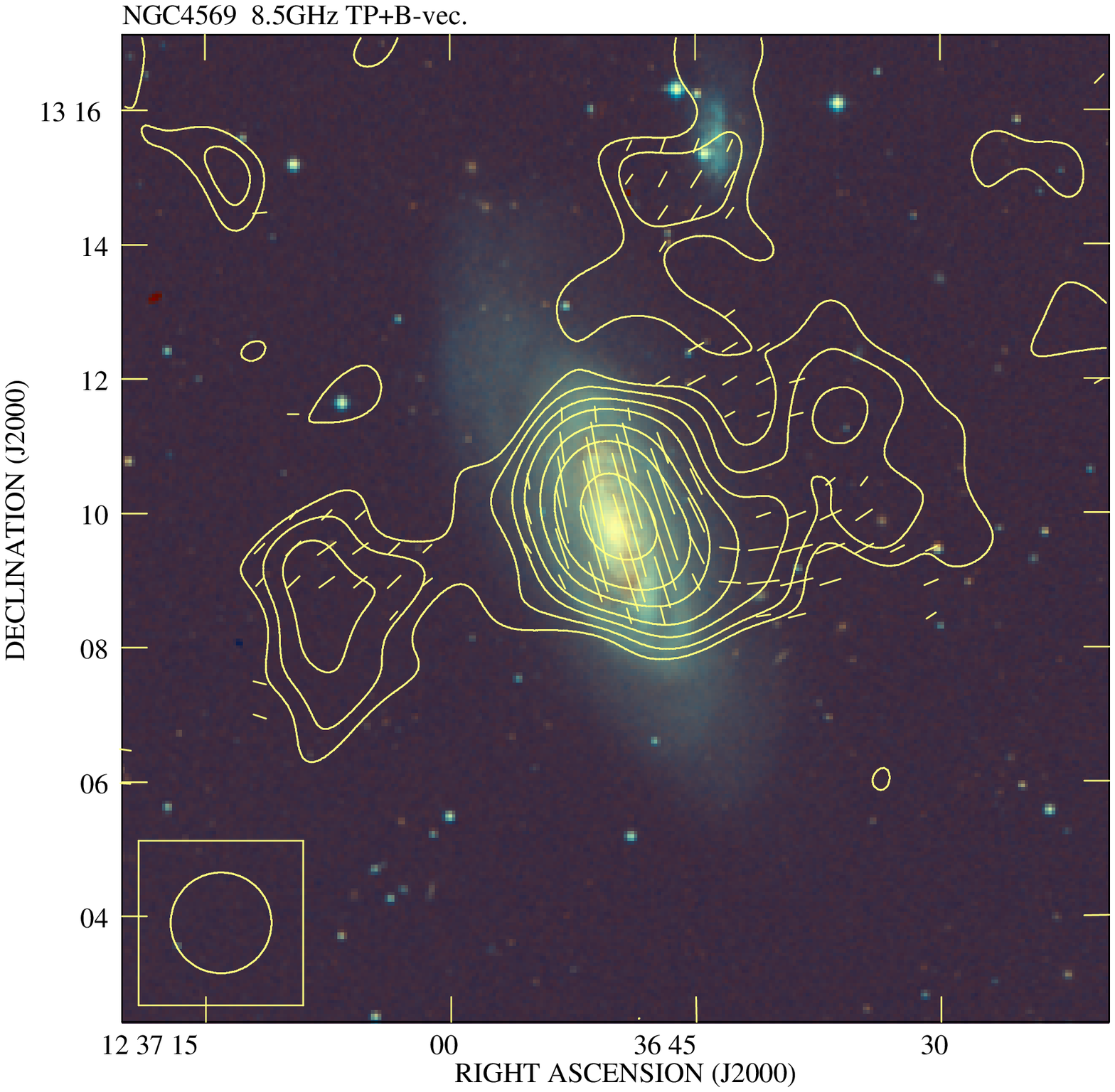}
\caption{Contours of total emission and polarization $B$-vectors of
the spiral galaxy NGC\,4569, observed at 8.4\,GHz and 90''\
resolution with the Effelsberg telescope. Vector lengths are
proportional to the polarized intensity (from Chy\.zy et al.
\cite{chyzy06}). } \label{fig:n4569}
\end{center}
\end{minipage}
\hspace{1cm}
\begin{minipage}[t]{6.5cm}
\begin{center}
\includegraphics[width=6cm]{stephan5}
\caption{Contours of the total power and $B$-vectors of Stephan's
Quintet observed with the VLA at 4.8\,GHz, overlaid upon an image
from the blue DSS. Vector lengths are proportional to the polarised
intensity (from Soida et al., in prep.).} \label{fig:stephan5}
\end{center}
\end{minipage}
\end{figure*}

Outside of the optical extent of a galaxy, little synchrotron
emission is detected at high frequencies because the highly
relativistic electrons responsible for the emission rapidly lose
energy as they travel away from the acceleration sites in supernova
remnants, which are concentrated in the inner disk of galaxies where
the star formation rate is highest. Radio halos are detected around
edge-on galaxies. The ``X-shaped'' structure observed in
polarization indicates action of a galactic wind (Krause
\cite{krause09}). In a few spectacular cases, huge radio lobes were
discovered, e.g. around NGC\,4569 which is located in the dense
Virgo cluster of galaxies (Fig.~\ref{fig:n4569}). Such phenomena
should be much more frequent at low frequencies.

Nearby starbursting dwarf galaxies are recognized for their poor
containment of magnetic fields and CRe (Klein et~al.
\cite{klein91}). This has led to the suggestion that they may
contribute considerably to the magnetization of the IGM (Bertone et
al. \cite{bertone06}), and may be responsible for generating
large-scale ordered magnetic fields through their disturbed
kinematics. LOFAR will be able to resolve the structure in nearby
dwarf galaxies out to distances of $\approx100$\,Mpc, where it will
be possible to detect CRe streaming into the extended halos and
place constraints on the generation mechanisms of the magnetic
fields in these regions.

Compression and shear can modify magnetic field structures. The
effects of these processes are most visible in tidally interacting
systems (e.g. the Antennae: Chy\.zy \& Beck \cite{chyzy04}) and in
galaxies interacting with intracluster gas (Vollmer et al.
\cite{vollmer10}). LOFAR will provide information on the strength
and structure of magnetic fields ejected into intergalactic space
during such interactions. In particular, observations of the Virgo
Cluster galaxies will enable extensive statistical studies of the
effects of ram pressure stripping by the intracluster gas and
high-velocity tidal interactions. This will yield conditions for
further MHD numerical simulations of the large-scale magnetic field
and polarized emission during galaxy evolution and for the 3-D
reconstruction of the intra-cluster magnetic field.

Galaxies at redshifts up to z\,$\approx$\,6 tend to cluster on the
scale of nearby groups (Conselice \cite{conselice07}); the
conditions in nearby compact groups may resemble those among field
galaxies at z\,$\approx$\,1--2. Consequently these groups constitute
unique laboratories for testing the evolution of galaxies and
intergalactic plasma. ``Magnetic enrichment'' of the IGM may be most
efficient in galaxy groups where bursts of star formation occurred
in the member galaxies. Diffuse synchrotron emission from colliding
galactic winds and amplified magnetic fields should be detectable
with LOFAR. Some groups even contain intergalactic gas pools (e.g.
Mulchaey et al. \cite{mulchaey03}) with large-scale shocks. Indeed,
recent observations of the groups HCG\,15 and Stephan's Quintet (SQ;
Fig.~\ref{fig:stephan5}) show the presence of partly ordered
intergalactic magnetic fields with an energy density comparable to
that of the thermal gas. The low frequencies provided by LOFAR will
be highly sensitive to such steep-spectrum shock-like features,
resembling relics in clusters, and knowledge of their 3-D magnetic
field structures from RM Synthesis will allow us a vastly improved
understanding of intergalactic gas dynamics: regular and anisotropic
fields produce clearly differentiated Faraday spectra. Low-frequency
studies may also reveal magnetized tails and cocoons with steep
radio spectra, providing essential information on how magnetic flux
field structure is supplied to intergalactic space beyond the group.
Improved knowledge of the intergalactic magnetic and CR pressure
affects the estimation of dark matter content within groups and
consequently the estimates of the mass parameter, $\Omega_{\rm{M}}$.

With the angular resolution and sensitivity of LOFAR, it will be
possible to obtain low-frequency maps providing completely new
information on the cosmic-ray electrons in nearby galaxies. The
cosmic-ray spectrum, derived from the radio synchrotron spectrum,
allows us to study and to understand the origin and propagation of
cosmic rays, the energy loss processes and how the propagation is
affected by the magnetic fields. Deep LOFAR observations of a sample
of spiral and dwarf galaxies are planned to observe diffuse
polarized emission and its Faraday rotation from the outer disks and
halos. LOFAR's sensitivity allows us to detect much fainter emission
than with present-day telescopes. An even more sensitive technique
to detect regular magnetic fields in galaxies is to observe a grid
of Faraday rotation measurements towards polarized background
sources. This method is independent of the presence of cosmic rays
in the galaxy and may allow us to detect regular fields at radial
and vertical distances larger than the detection limit of the
synchrotron emission. The MKSP Project aims to clarify the origin of
magnetic fields in galactic halos. Proposed models are dynamos which
can generate large-scale regular fields, or galactic winds where
magnetic fields from the disk are blown out and amplified by the
outflow.

First LOFAR maps for M~51 and NGC~4631 are presented by Mulcahy et
al. (this volume).

\subsection{Intergalactic filaments}

The search for magnetic fields in the intergalactic medium (IGM) is
of fundamental importance in cosmology.
The prediction of a large-scale ``Cosmic Web'' is one of the
defining characteristics of large-scale structure simulations. The
warm--hot intergalactic medium (WHIM) contained in this web may
account for the missing two thirds of baryon density in the Universe
expected from concordance cosmology and studying the largely unknown
nature of the magnetic fields in these environments is essential for
understanding the origin and evolution of magnetism in the Universe.
With LOFAR it will be possible to search for synchrotron radiation
from the Cosmic Web at the lowest possible levels. Such emission
probes the existence of magnetic fields in the most rarefied regions
of the IGM, and measuring their intensity provides a means of
investigating both their origin and relation to large-scale
structure formation in the Universe. Faint emission has previously
been detected around the Coma cluster (Kronberg et~al.
\cite{kronberg07}), although it is debated whether this emission can
truly be attributed to filamentary plasma. Deep LOFAR observations
of Coma, sensitive to a wide range of angular scales, will be
crucial for confirming the current data, and for definitively
detecting ultra-steep-spectrum emission associated with an aging
relativistic electron population.

\begin{figure}[th]
\begin{center}
\includegraphics[width=9cm]{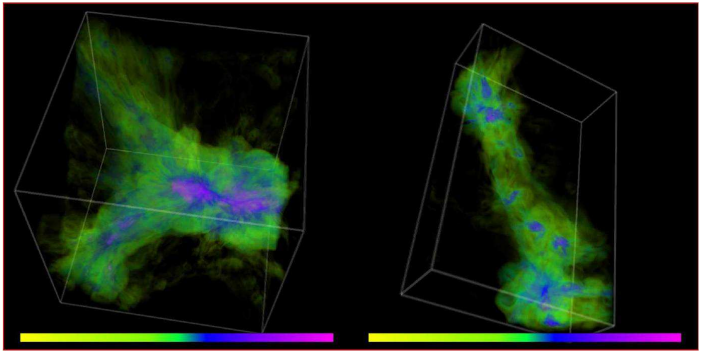}
\end{center}
\caption{Simulation of magnetic fields in the Cosmic Web at $z = 0$
in a volume of $25\, h^{-1}$\,Mpc$^3$ centered around a cluster
complex (left panel) and in a volume of $25 \times 15.6 \times
6.25\,h^{-1}$\,Mpc$^3$ which includes a number of galaxy groups
along a filament (right panel). The color codes the magnetic field
strength (logarithmically scaled) from 0.1\,nG (yellow) to 10~$\mu$G
(magenta). Clusters and groups are shown with magenta and blue,
while filaments are green (from Ryu et al. \cite{ryu08}). }
\label{fig:ryu}
\end{figure}

LOFAR is well suited for the search for intergalactic magnetic
fields. If the Cosmic Web outside clusters also contains a magnetic
field (Fig.~\ref{fig:ryu}) we can hope to detect this field by
either direct observation of synchrotron emission or the Faraday
rotation of polarized emission from background sources. Detection of
this field, or placing stringent upper limits on it, will provide
powerful observational constraints on the origin of cosmic
magnetism. The high Faraday depth resolution provided by the broad
wavelength-squared coverage of LOFAR will also allow detection of
weak magnetic fields in the cosmic web using RMs. Akahori \& Ryu
({\cite{akahori11}) predict that the variance in the RM due to
intergalactic magnetic fields will be of order 1\,rad/m$^2$, meaning
that precise RM measurements, with spectral filtering to take
account of self-absorption polarization effects, are required if one
wants to separate the contribution of cosmic filaments from stronger
intervening sources (e.g. the Milky Way foreground, see Sect.~2.1).
A statistical analysis is needed like the measurement of the power
spectrum of the magnetic field of the Cosmic Web (Kolatt
\cite{kolatt98}) or the cross-correlation with other large-scale
structure indicators like the galaxy density field.

\subsection{Giant radio galaxies}

Giant radio galaxies (GRGs), defined as objects with dimensions
larger than about 1\,Mpc, are the largest single objects in the
Universe (see e.g. Machaski et al. \cite{machalski06}) and are
extremely useful for studying a number of astrophysical problems.
These range from understanding the evolution of radio sources,
constraining the orientation-dependent unified scheme for AGN, to
probing the intergalactic and intercluster medium at different
redshifts.

The low-energy electrons responsible for the low-frequency emission
can propagate large distances from their origins in the central core
or lobe hotspots, and large radio cocoons are expected to surround
many objects. Using high angular resolution observations with the
full international LOFAR array should help us to learn more about
the low-energy electron population in these objects, and hopefully
to better understand the acceleration mechanisms which produce the
relativistic electrons responsible for synchrotron emission.

Because of the low density in and around these giant sources, with
lobes located far outside the gaseous spheres of the parent
galaxies, GRGs can be expected to be highly polarized even at low
frequencies. The high degree of polarization of giant radio galaxies
also makes them ideal polarization calibrators for the observation
of weaker sources.

\subsection{Stellar jets}

%

Jets from young stars are one of the most striking manifestations of
star formation. A crucial part of the accretion/ejection mechanism,
they are ubiquitous across low and high mass star formation.
Emission indicative of high magnetic field strengths, attributed to
both synchrotron (Carrasco-Gonz{\'a}lez et~al. 2011) and
gyrosynchrotron (Ray et~al. 1997, Scaife et~al. 2011) has been seen
from a number of objects. We plan to use spatially resolved
polarized structure (linear \& circular) in these jets to examine
the magnetic field structure of outflows and to investigate the
impact of magnetic fields on the launching and evolution of
protostellar jets. The MKSP plans to use the large FoV of LOFAR to
observe multiple objects simultaneously by targeting three regions
with a high density of star formation: the Taurus, Perseus \&
Cepheus Flare (CF) molecular clouds, at sub-arcsecond resolution.
These regions are selected to be nearby ($\leq$~300 pc) to allow
good physical resolution and to provide contrasting samples of
protostars in different stages of evolution.

\section{MKSP observing program}


The core of the MKSP is to deeply map in the LOFAR highband
(120--180\,MHz) the diffuse total and polarized emission and its
Faraday rotation (RM) distribution from selected regions in the
Milky Way, a variety of nearby galaxies, selected galaxy groups, and
the galaxies of the Virgo cluster. The diffuse total emission from
extended disks and halos around nearby galaxies is best detected at
low resolutions of 10''--60''.


Sub-areas of the deep fields centered on polarized background
sources will be imaged with high resolution (about 1'') to obtain RM
grids. LOFAR presently has 8 international stations with baselines
up to about 1000\,km, providing sub-arcsecond resolution (Heald et
al. \cite{heald11}). RM grids are the most sensitive way to detect
weak regular fields because the signal-to-noise ratios are much
higher than that of the diffuse galactic emission.
A minimum of about 10 background sources is needed to recognize a
large-scale field pattern in a galaxy (Stepanov et al.
\cite{stepanov08}). High angular resolution is needed here, too. The
deep fields around galaxies will be located at different Galactic
latitudes and will also be analyzed with respect to the properties
of the small-scale magnetic field in the foreground of the Milky
Way, e.g. by computing the structure functions as a function of
Galactic latitude.

Detecting RM signals from intergalactic magnetic fields is a
challenge which requires a very large areal source density and hence
a very high sensitivity. Studies of diffuse emission from
intergalactic filaments (Sect.~2.4) will use the SKSP Tier~1 and
Tier~2 surveys that include the Coma field. Deeper follow-up
investigations will be made with the Tier~2 survey. These
investigations will focus on a $\approx$\,100 sq.~degree area
centered on the Coma cluster of galaxies. Proof for an intergalactic
origin of (part of) the RM could come from a statistical comparison
with source redshift. The best candidates are deep fields around
compact galaxy groups with minimum Galactic foreground contribution.

Furthermore, polarization of selected giant radio galaxies will be
observed in three frequency bands, and polarization of a few stellar
jets in the highband.


\section{Polarization calibration}

Polarization calibration for LOFAR is mostly a matter of dealing
with the ionosphere. The typical Faraday rotation introduced by the
Earth's ionosphere is 1--3\,rad/m$^2$. In order to calibrate the
angle of linear polarization to within $10^\circ$ on the sky at
120\,MHz, the ionospheric Faraday rotation must be corrected to
better than 1\% of its absolute value for each LOFAR station. As
polarized emission is significantly weaker than total emission, and
as there are few strongly polarized sources for use as calibrators,
ionospheric Faraday rotation calibration is a challenge for LOFAR.

Many observing plans for LOFAR polarimetry need the wide field of
view to simultaneously measure hundreds of sources, or to measure
large sources in single pointings. These observations require
accurate polarization calibration across the entire LOFAR beam,
including the individual antenna beams, the high band tile beams,
the phased-array station beams, and potentially tied-array beams.
Science targets of interest often have polarized emission a factor
of 10 or 100 times weaker than the total emission. Hence, the
polarization calibration of the leakage terms of the LOFAR beams
must be accurate to better than one percent in order to prevent
leakage of the total intensity into polarization intensity through
mis-calibration from dominating the real signal.

%
%

\section{RM Synthesis}

Successful low frequency observations need digital backends with
numerous narrow-frequency channels to cope with man-made
interference signals and to reduce depolarization by Faraday
rotation across the observing band. Multi-channel polarization data
are used for the technique of {\em Rotation Measure Synthesis}. It
Fourier-transforms multifrequency polarization data into a data cube
with Faraday depth as the third coordinate (Brentjens \& de Bruyn
\cite{brentjens05}, Heald \cite{heald09}). The Faraday depth (FD) is
the integral of the plasma density times the field strength along
the line of sight. Information about the distribution of regular
magnetic fields and ionized gas along lines of sight is obtained
and, with help of some modeling, a 3-D picture of cosmic magnetic
fields can be derived (Bell et al. \cite{bell11a}, Frick et al.
\cite{frick11}).


LOFAR RM data cubes will have Faraday depth dimensions of perhaps
tens of thousands of pixels. A full RM cube from a single pointing
with international LOFAR baselines would be well over 100\,Petabytes
in size. Data processing will have to be performed in faceted chunks
for many years. Adding to the computational challenge is the need to
deconvolve (``CLEAN'') the RM cube in Faraday depth space as well as
position space (Heald \cite{heald09}). The MKSP Team is developing
single-dimensional deconvolution algorithms for LOFAR, with 3-D
deconvolution methods planned for future research (Bell \&
En{\ss}lin \cite{bell11b}).


LOFAR observations with large signal-to-noise in the highband
(120--180\,MHz) will allow the detection of FD with precisions well
below 1\,rad/m$^2$. This is sufficient to detect fields below
1\,$\mu$G (0.1\,nT), which has never been possible to date. In the
LOFAR lowband range (15--70\,MHz), the FD accuracy is formally even
better, but the calibration is more difficult and the sensitivity is
lower than in the highband range.

The first successful LOFAR detection of polarized emission and RM
from the pulsar J0218+4232 was presented by Heald et al.
(\cite{heald11}).

\section{Project Team}

At the beginning of 2012, the Project Team consists of 29 full and
54 associated members from 14 countries. The Project is led by a
German/Dutch/UK management team.

MKSP website:
http://www.mpifr-bonn.mpg.de/staff/rbeck/MKSP/mksp.html

\acknowledgements{LOFAR, the Low Frequency Array designed and
constructed by ASTRON, has facilities in several countries, that are
owned by various parties (each with their own funding sources), and
that are collectively operated by the International LOFAR Telescope
(ILT) foundation under a joint scientific policy. -- This research
is supported by the DFG Research Unit 1254.}

\end{document}